\documentclass[aps,twocolumn,groupedaddress]{revtex4}
\usepackage{amsfonts}
\usepackage{amssymb}
\usepackage[cp850]{inputenc}
\usepackage{times}
\usepackage{mathptmx}
\usepackage{graphicx}
\usepackage{epsf}
\begin{document}
\bibliographystyle{apsrev}
\title{Probing ion transport at the nanoscale: Time-domain electrostatic force spectroscopy on glassy electrolytes}
\author{A. Schirmeisen, A. Taskiran, H. Fuchs}
\affiliation{Physikalisches Institut and Center for Nanotechnology (CeNTech), Westfälische Wilhelms-Universität Münster, 
Wilhelm-Klemm-Str. 10, 48149 Münster, Germany}
\author{B. Roling, S. Murugavel} \affiliation{Institut für Physikalische
Chemie and CeNTech , Westfälische Wilhelms-Universität Münster, Corrensstr. 30, 48149 Münster, Germany}
 \author{H. Bracht, F. Natrup}\affiliation{Institut für Materialphysik and CeNTech, Westfälische 
Wilhelms-Universität Münster, Wilhelm-Klemm-Str. 10, 48149 Münster, Germany}
\date{\today}
\begin{abstract}
We have carried out time--domain electrostatic force spectroscopy on two different ion conducting glasses using
an atomic force microscope. We compare the electrostatic force spectroscopic data obtained at different temperatures 
with macroscopic electrical data of the glasses. The overall consistency of the data shows that electrostatic 
force spectroscopy is capable of probing the ion dynamics and transport in nanoscopic subvolumes of the samples.
\end{abstract}

\maketitle

Ion conducting crystals, glasses and polymers are widely used as solid electrolytes in batteries, fuel cells, and
chemical sensors. A lot of research work is being carried out in order to find new materials with improved ionic
conductivities. One method that is becoming more and more technologically relevant is nanostructuring of materials. It has,
for instance, been found that the ionic conductivity of nanocrystalline ionic conductors can be increased by adding
nanocrystalline insulators \cite{Indris00}. In the case of glasses, a conductivity enhancement can be achieved by the
formation of nanocrystallites during partial crystallisation \cite{Adams96}. Furthermore, the ionic
conductivity of polymer electrolytes can be improved considerably by incorporating nanoparticles, such as Al$_2$O$_3$,
TiO$_2$ and ZrO$_2$, into the polymer matrix \cite{Scrosati00}.\par
Up to now, there is no general agreement about the origin of these conductivity enhancement effects. A limiting factor 
hindering a better theoretical understanding and thus a more systematic preparation of improved materials
is the traditional characterization of the ion dynamics by means of macroscopic techniques, such as conductivity
spectroscopy, tracer diffusion measurements, and NMR relaxation techniques. In nanostructured solid electrolytes, 
diffusion pathways in different phases and at interfaces are believed to play an important role for the ion transport 
\cite{Indris00, Maier84}. Therefore, an experimental method capable of probing ion transport on nanometer length scales 
would be highly desirable.\par
In principle, electrostatic force microscopy and spectroscopy techniques using an atomic force microscope (AFM) are well 
suited for this purpose. Such techniques
have been applied in different research fields to characterize the electrical properties of materials on nanoscopic length
scales. Scanning capacitance microscopy \cite{Zavyalov99,Brezna02,Giannazzo02} and scanning kelvin probe microscopy
\cite{Baikie98,Jacobs98} have been applied to semiconductors and semiconductor devices. The electrical properties of
nano\-struc\-tured materials adsorbed on insulating substrates, e.g. carbon nanotubes and DNA molecules adsorbed on silica,
have been studied by using electrostatic force microscopy techniques \cite{Gil02,Moreno03}. Israeloff and coworkers 
used time--domain electrostatic force spectroscopy in order to characterize dielectric fluctuations in thin 
polymer films at the glass transition \cite{Walther98, Russell98, Russell00}. \par
In this letter, we report, for the first time, on the application of electrostatic force 
spectroscopy for studying ion transport in solid electrolytes. To do this, we chose two 
ion conducting glasses with well--known macroscopic electrical properties. The chemical
compositions of these glasses are 0.25 Na$_2$O $\cdot$ 0.75 GeO$_2$ (NG glass) and 0.143 K$_2$O $\cdot$ 0.286 CaO
$\cdot$ 0.571 SiO$_2$ (KCS glass). The activation energies of the dc electrical conductivity, $E_A^{dc}$, reflecting the
thermally activated long-range alkali ion transport, are 0.74~eV in the case of the NG glass and 1.05~eV in the case of
the KCS glass, respectively. This implies that for a given temperature, the Na$^+$ ions in the NG glass are much more
mobile than the K$^+$ ions in the KCS glass. Since the structure of both glasses is homogeneous on length scales of
10-30~nm, the differences in the mobility of the alkali ions should also manifest in the nanoscopic electrical
properties as probed by electrostatic force spectroscopy.\par
For the preparation of the KCS and NG glasses special attention was paid to the surface quality. Samples of typically
4x8 mm$^2$ lateral dimensions and 2~mm thickness were cut from the respective bulk glasses. One surface of each sample
was first mechanically grinded with sandpaper of 15 micron grain size. Subsequently, a four-step polishing procedure
was carried out with diamond suspensions of various grades. The first and second steps were performed on a silk pad with
aqueous diamond suspensions of 9 and 6 micron grain size, respectively. Then a polishing with 1 micron grade and
finally with a water-free 1/4 micron suspension was conducted on a felt pad. In this way we achieved a scratch- and
pit-free surface with an overall roughness of less than 4~nm.\par
\begin{center}
\begin{figure}[htb]
\begin{center}
\epsfxsize=8cm \leavevmode{\epsffile{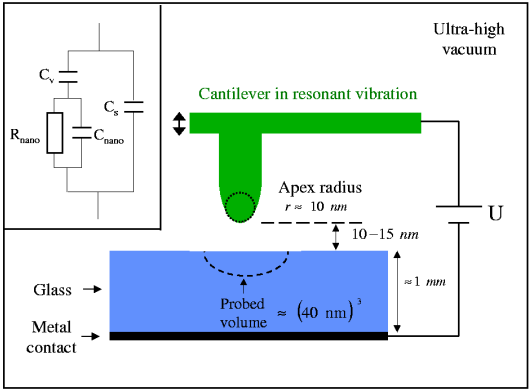}} 
\caption{Schematic illustration of the experimental setup for electrostatic 
force spectroscopy on ion conducting glasses. The equivalent circuit models the 
overall capacitance of the system between tip and metal contact.}
\label{setup} 
\end{center}
\end{figure}
\end{center}

For the electrostatic force spectroscopy we use a commercial, variable temperature AFM operating under ultrahigh
vacuum (UHV) conditions (Omicron VT-AFM). The force sensor is a single crystalline, highly doped silicon cantilever
with a resonant frequency $f_0$ = 300 kHz and a spring constant $k$ = 20~N/m. The apex radius of the sharp conducting 
tip is 10~nm (non-contact cantilever NCHR from Nanosensors). A schematic diagram of the setup is shown in Fig.~\ref{setup}. 
While the tip can be biased with a voltage $U$ up to 10 V, the glass samples are attached to a grounded metal contact.
The system is operated in the frequency modulation mode (FM mode) \cite{Albrecht91},
where the cantilever is oscillated at its resonant frequency. Conservative tip-sample forces will induce a shift in the
resonant frequency of the cantilever, which is used as the feedback parameter for the tip-sample distance control. 
In the case of small cantilever oscillations and long ranged electrostatic interactions, the frequency shift $\Delta f$ 
induced by the bias voltage $U$ is given by \cite{Walther98}:
\begin{equation}
\label{eq_f_C}
\Delta f(t) \;=\; - \, \frac{f_0}{4\,k} \cdot U^2 \cdot \frac{d^2 C(t)}{d\,z^2}
\;,
\end{equation}
where $C(t)$ denotes the overall capacitance between biased tip and ground. $C(t)$ can be modeled by the equivalent circuit 
illustrated in Fig.~\ref{setup}. The probed nanoscopic subvolume of the sample is represented by a resistor $R_{nano}$ 
in parallel to a capacitor $C_{nano}$. The gap between tip and the probed sample volume is modeled by a vacuum capacitor 
$C_V$ in series to the $R_{nano} \, C_{nano}$ element. Additionally we introduce a capacitor $C_S$ in parallel to the 
other elements, which represents all stray capacitances between tip and ground. When the voltage 
$U$ is applied, all capacitors are instantaneously charged. Subsequently, the capacitor $C_{nano}$ is discharged 
through the resistor $R_{nano}$. This leads to an increase of the overall capacitance $C(t)$ and thus to a decrease of the 
resonant frequency $f(t)$. The time dependence of $C(t)$ is given by:
\begin{equation}
\label{eq_equiv_circuit}
C(t) \;=\; C_V \left[ 1 \,-\,\frac{C_V}{C_{nano} + C_V} \cdot \exp(-t/\tau)
\right] \,+\, C_S
\end{equation}
with $\tau = R_{nano} \cdot (C_{nano} + C_V)$. From a microscopic point of view, the discharge of the sample capacitor 
is due to mobile ions moving in the direction of the electric field, until the field in the probed volume becomes zero. Thus, the 
time dependent change of the cantilever resonance frequency reflects the time dependent motion of the mobile ions.\par 
An important feature of this measurement technique is the sample volume which is probed with the nanoscopic tip.
Detailed finite element simulations using the FEMLAB software yield an approximate probed sample volume of (40 nm)$^3$. 
Taking into account the number densities of the alkali ions in our glass samples of
$N_V$(Na$^+$) = 1.21$\cdot 10^{22}$~cm$^{-3}$ (NG glass) and $N_V($K$^+$) = 7.2$\cdot 10^{21}$~cm$^{-3}$ (KCS glass) we expect 
to measure the dynamic behaviour of an ensemble of less than 10$^6$ ions, many orders of magnitude below the macroscopic 
level.\par
For the time--domain spectroscopic measurements, we apply the following procedure: The
tip is approached towards the sample at zero tip voltage. In this case, mainly attractive van der Waals forces are
acting between tip and sample surface leading to a negative frequency shift of the cantilever resonance. With an
oscillation amplitude of 4~nm, a tip radius of 10~nm and a frequency shift of -20~Hz, the tip-sample distance is
typically in the range of 5 to 10~nm. In the FM mode the surface is scanned to ensure that the following spectroscopic
measurements are performed on a flat and uniform part of the surface. The tip is then
positioned, and the distance-feedback mechanism is disabled. The tip is retracted by 10~nm and then instantaneously 
biased with $U$ = -4~V. The dynamic processes in the probed volume are monitored by measuring the frequency shift as a 
function of time until the saturation value is reached. Our system is capable to reliably measure saturation
times from 0.001~s up to 10~s, covering 4 orders of magnitude. The main complication during these
measurements is temperature--induced drift of the tip-sample distance, especially at elevated temperatures. We developed
a technique where the temperature drift rate is measured by the AFM, before the voltage is applied. Based on the assumption 
that the drift rate is constant over a period of a few minutes, we compensate for the drift during
electrostatic force measurements. \par
\begin{center}
\begin{figure}[htb]
\begin{center}
\epsfxsize=8cm\leavevmode{\epsffile{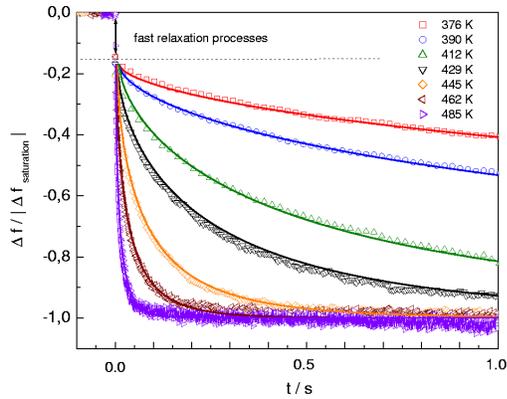}}
\caption{Time dependence of the cantilever resonance frequency
$\Delta\,f$ after applying a voltage of $U = -4 V$ to the
0.143 K$_2$O $\cdot$ 0.286 CaO $\cdot$ 0.571 SiO$_2$ glass.
The frequency axis is normalised by the absolute value of
the maximal frequency shift, $\mid \Delta\,f_{\rm saturation} \mid$.
The solid lines represent best fits of the experimental data
to the stretched exponential function given by Eq. (\ref{eq_kww}).}
\label{relax_curves}
\end{center}
\end{figure}
\end{center}
\begin{center}
\begin{figure}[htb]
\begin{center}
\epsfxsize=8cm \leavevmode{\epsffile{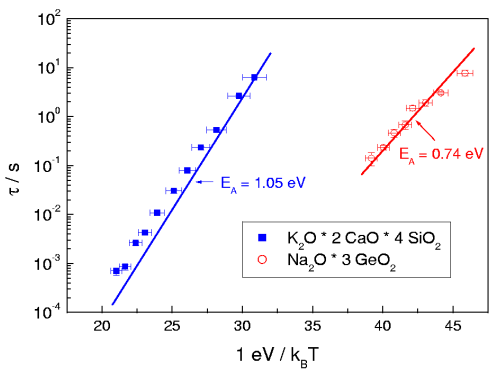}} 
\caption{Arrhenius plot of the nanoscopic relaxation times $\tau$ (symbols)
and of the macroscopic relaxation times $\tau_{macro} = 
R_{macro}\cdot C_{macro}$ (solid lines).}
\label{arrh}
\end{center}
\end{figure}
\end{center}

In the case of the KCS glass, the time dependence of $\Delta f$ was measured at different temperatures in a range from 
376~K to 570~K. Fig.~\ref{relax_curves} shows selected relaxation curves. 
These curves were normalized to their respective saturation 
frequency shift values $\Delta f_{\rm saturation}$ (which were in the range -200 Hz $\pm$ 20 Hz) for a better comparison 
of the relevant time scales. The relaxation curves were fitted with a stretched exponential function of the form 
\cite{Walther98}:
\begin{equation}
\label{eq_kww}
\Delta\,f(t) \;=\; (\Delta\,f_{\rm saturation} - \Delta\,f_{\rm fast}) \cdot \left[ 1\,-\,\exp(-(t/\tau)^{\beta})
\right]
+ \Delta\,f_{\rm fast}
\end{equation}
Here, $\Delta\,f_{\rm fast}$ denotes the frequency shift due to fast relaxation processes
(vibrational and electronic polarization).
The same type of measurement was also performed on the NG glass sample in a temperature range from
253 K to 296 K. The stretch factors are $\beta$ = 0.65 $\pm$ 0.05 for the
KCS glass and $\beta$ = 0.55 $\pm$ 0.05 for the NG glass, respectively. In Fig.~\ref{arrh} we plot the logarithm of the 
relaxation time $\tau$ versus the inverse temperature. 
The data of both samples follow, to a good approximation, an Arrhenius 
law. This is expected since the resistance $R_{nano} = \tau / (C_{nano}+C_V)$ is determined by the thermally activated ion 
tranport in the probed subvolume.

The solid lines in Fig.~\ref{arrh} represent the macroscopic relaxation times $\tau_{macro} = R_{macro}\cdot C_{macro}$,
which we calculated from the macroscopic resistances $R_{macro}$ and capacitances $C_{macro}$ of the samples.
Clearly, there is good agreement between the relaxation times $\tau$ und $\tau_{macro}$ and their
respective temperature dependences. This shows that the same dynamic processes are probed by electrostatic force spectroscopy and by macroscopic 
electrical spectroscopy, namely the dynamics and transport of the mobile ions. However, a
quantitative comparison between 
$\tau$ and $\tau_{macro}$ is difficult for two reasons. First, the values of $\tau$ are influenced by the vacuum capacitor 
$C_V$. Second, the description of the electrical properties of the nanoscopic sample volume by a parallel RC element is 
an approximation. The more complex electrical properties of both glass samples lead to a {\it stretched} exponential 
relaxation of the experimental $C(t)$ data as compared to the simple exponential time dependence of $C(t)$ in
Eq. (\ref{eq_equiv_circuit}).

In conclusion, we have shown that electrostatic force spectroscopy can be used to probe the ion dynamics and transport in 
nanoscopic subvolumes of solid electrolytes. We note that the high sensitivity of this method opens up the possibility 
to study discrete dynamic events, i.e. hopping motions of {\it single ions} between neighboring potential
minima. Furthermore, we emphasize that electrostatic force spectroscopy should be a very powerful method for studying 
the dynamics in nanostructured solid electrolytes and other nanostructured materials.


\end{document}